\documentclass[journal]{IEEEtran}

\ifCLASSINFOpdf
\else
   \usepackage[dvips]{graphicx}
\fi
\usepackage{url}

\usepackage{graphicx}
\usepackage{subfigure}
\usepackage{eqparbox}
\usepackage{algorithm}
\usepackage{algorithmicx}
\usepackage{algpseudocode}
\usepackage{amsmath}
\usepackage{booktabs}
\usepackage[colorlinks,linkcolor=black,anchorcolor=black,citecolor=black]{hyperref}

\begin{document}

\title{FFT-Enhanced Low-Complexity Near-Field Super-Resolution Sensing}
\author{Yuxiao Wu, Huizhi Wang, \IEEEmembership{Graduate Student Member, IEEE} and Yong Zeng, \IEEEmembership{Senior Member, IEEE}
\thanks{ This work was supported by the Natural Science Foundation for Distinguished Young Scholars of Jiangsu Province with grant number BK20240070. (Corresponding author: Yong Zeng.)}
\thanks{Yuxiao Wu and Huizhi Wang are with the National Mobile Communications Research Laboratory, Southeast University, Nanjing 210096, China (e-mail: 220240864@seu.edu.cn; wanghuizhi@seu.edu.cn).}
\thanks{Yong Zeng is with the National Mobile Communications Research Laboratory, Southeast University, Nanjing 210096, China, and also with the Purple Mountain Laboratories, Nanjing 211111, China (e-mail: yong\_zeng@seu.edu.cn).}}

\maketitle

\begin{abstract}
In this letter, a fast Fourier transform (FFT)-enhanced low-complexity super-resolution sensing algorithm for near-field source localization with both angle and range estimation is proposed. Most traditional near-field source localization algorithms suffer from excessive computational complexity or incompatibility with existing array architectures. To address such issues, this letter proposes a novel near-field sensing algorithm that combines coarse and fine granularity of spectrum peak search. Specifically, a spectral pattern in the angle domain is first constructed using FFT to identify potential angles where sources are present. Afterwards, a 1D beamforming is performed in the distance domain to obtain potential distance regions. Finally, a refined 2D multiple signal classification (MUSIC) is conducted within each narrowed angle-distance region to estimate the precise location of the sources. Numerical results demonstrate that the proposed algorithm can significantly reduce the computational complexity of 2D spectrum peak searches and achieve target localization with high-resolution.
\end{abstract}


\IEEEpeerreviewmaketitle

\section{Introduction}

\IEEEPARstart{E}{xtremely} large-scale MIMO (XL-MIMO) has emerged as a promising technique to significantly enhance spectral efficiency and spatial resolution in 6G \cite{XL-MIMO}.
However, the evolution from 5G massive MIMO to 6G XL-MIMO fundamentally alters channel characteristics. Sources are no longer absolutely located in the far-field region, but are likely to be in the near-field region with a spherical wavefront model \cite{TSP}.
Consequently, classical super-resolution estimation algorithms for far-field source localization are no longer directly applicable. Note that the term “super-resolution” means a resolution exceeding that achieved by the DFT-based algorithms limited by the array aperture, bandwidth, or duration of the signal \cite{super-resolution}. Typical super-resolution algorithms include the Multiple Signal Classification (MUSIC) and the Estimation of Signal Parameters via Rotational Invariance Techniques (ESPRIT). 

 Various near-field source localization algorithms have been proposed. For example, the MUSIC algorithm is extended for near-field source localization in \cite{2D-MUSIC}, achieving joint estimation of the direction of arrival (DoA) and distance of near-field sources. This method can achieve good estimation performance, but the angle-distance 2D peak search significantly increases the computational complexity. 
 Two low-complexity algorithms based on second-order statistics (SOS) are proposed in \cite{2-order1} and \cite{2-order2}, but they attain poor estimation performance. To reduce the complexity while maintaining high resolution, a reduced-rank algorithm based on the MUSIC algorithm (RR-MUSIC) is proposed in \cite{RR-MUSIC}, which decouples the angle and distance parameters and transforms the 2D peak search into multiple 1D peak searches. Furthermore, a reduce-dimension MUSIC algorithm (RD-MUSIC) is proposed in \cite{RD-MUSIC}, which requires only one single 1D peak search, further reducing complexity. However, both algorithms in \cite{RR-MUSIC} and \cite{RD-MUSIC} require antenna spacing to be no greater than one quarter of the signal wavelength, which is incompatible with current array architectures \cite{sparse-MIMO}. In the field of near-field channel estimation, \cite{two-phase} proposes a two-phase beam training method that decomposes the 2D search into two sequential phases, which utilizes conventional far-field codebooks to obtain the candidate angles and then a customized polar-domain codebook to find the precise distance.
 Beyond merely decoupling the angle and range parameters, both \cite{FFT-MUSIC1} and \cite{FFT-MUSIC2} utilize the fast Fourier transform (FFT) algorithm for an initial coarse estimation, which is followed by the MUSIC algorithm within the narrowed estimation range for high-resolution estimation. 
 However, neither \cite{FFT-MUSIC1} nor \cite{FFT-MUSIC2} has taken into account the near-field scenario. 
 
This letter proposes a novel super-resolution algorithm for near-field localization based on FFT, without extra requirement for inter-element spacing, making it compatible with typical array architectures. The proposed algorithm employs the FFT algorithm to first exclude angle regions without any source. Then, the candidate angle regions are further estimated in distance domain using far-field beamforming, which will effectively narrow the search range, thus significantly reducing the complexity of the 2D-MUSIC algorithm, yet achieves comparable estimation performance. 

\section{system model}

\begin{figure}[htbp] 
\centering
\includegraphics[width=0.45\hsize]{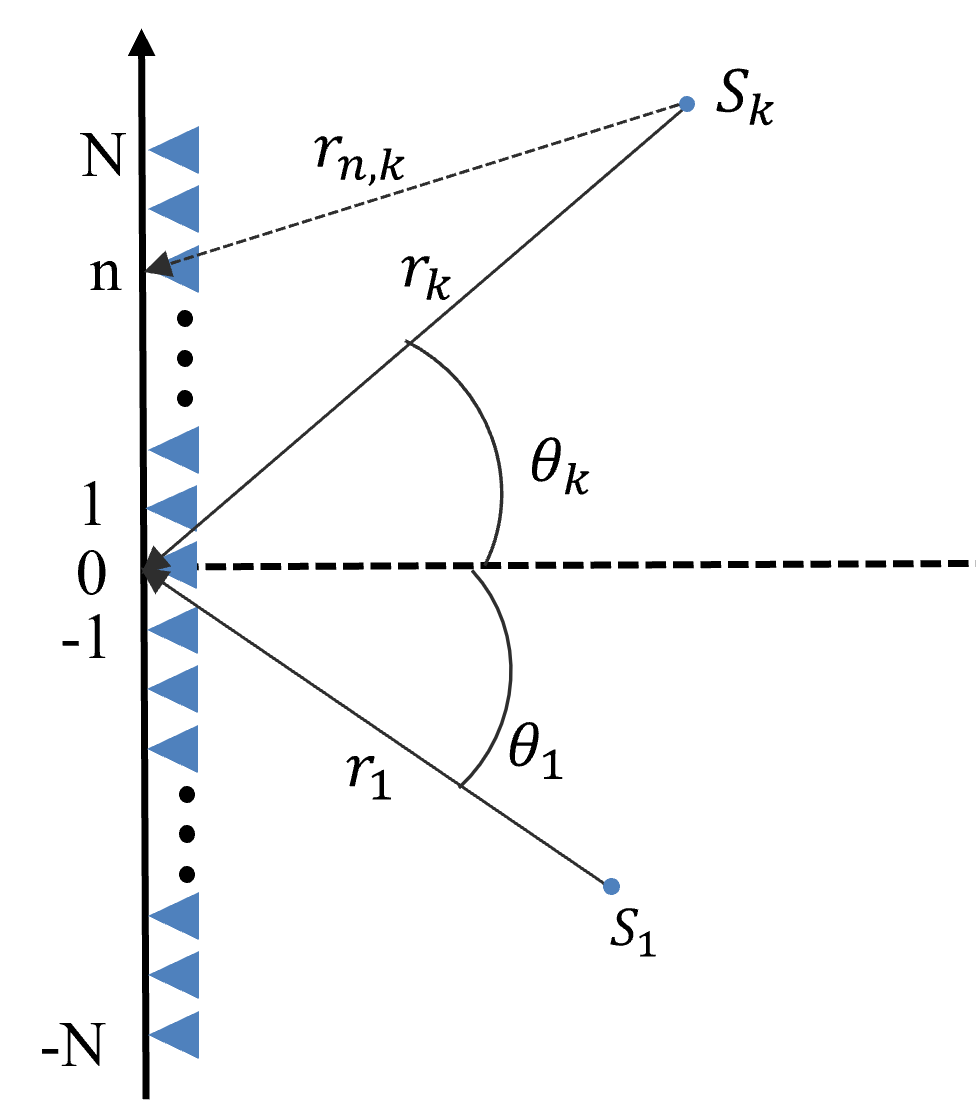}
\caption{ System model for near-field source localization. }
\label{fig:1}
\end{figure}

As shown in Fig.\ref{fig:1}, the base station (BS) is equipped with an extremely large-scale uniform linear array (ULA) consisting of $M=2N+1$ elements, which is used to receive signals emitted by signal sources located at $K$ distinct positions. The inter-element spacing of the ULA is denoted by $d$, and the index of each antenna element is represented by $\delta_n =-N, \dots, N.$ The  coordinate of each antenna element is $\delta_n d$, with the antenna element $\delta_n =0$ designated as the reference element. Distance and DoA from the $k$-th source to the reference element are denoted by ($r_k,\theta_k$) for $k = 1,2,
\dots,K$, where $r_k \in [r_{\rm{\min }},r_{\rm{\max }}]$ and $\theta_k \in [\theta_{\rm{\min }},\theta_{\rm{\max }}]$.

The distance from source $k$ to the $\delta_n$-th array element is 
\begin{equation}
    r_{n,k}=\sqrt{(\delta_nd)^2+r_k^2-2\delta_ndr_k\sin\theta_k}. \label{eq:2.1}
\end{equation}
The corresponding phase shift of the $\delta_n$-th element relative to the reference element is 
\begin{equation}
    a_n(r_k,\theta_k)=e^{-j\frac{2\pi}{\lambda}(r_{n,k}-r_k)}, \label{eq:2.2}
\end{equation}
where $\lambda$ is the wavelength of the signal.
The steering matrix $\boldsymbol{A}$ is then given by
\begin{equation}
    \boldsymbol{A}=[\boldsymbol{a}_1,\dots,\boldsymbol{a}_K], \label{eq:2.3}
\end{equation}
\vspace{-1.4em}
where $\boldsymbol{a}_k=[a_{-N}(r_k,\theta_k),...,1,  ...,a_N(r_k,\theta_k)]^T$.
Thus, the received signal for the entire antenna array is
\begin{equation}
\boldsymbol{y}(i)=\boldsymbol{A}\boldsymbol{x}(i)+\boldsymbol{z}(i), i=1,\dots,J,
    \label{eq:2.4}
\end{equation}
where $\boldsymbol{x}(i)$ represents the transmitted signals that are independently distributed, with the covariance matrix $E[\boldsymbol{x}(i)\boldsymbol{x}(i)^H]=\boldsymbol{R}_s$ being a diagonal matrix, $J$ represents the number of snapshots, and $\boldsymbol{z}$ is the additive white Gaussian noise (AWGN) with variance $\sigma^2$ and mean zero. Thus, the received signal matrix can be represented as $\boldsymbol{Y} = [\boldsymbol{y}(1), \ldots, \boldsymbol{y}(J)]$. 
For near-field source localization, it is hoped that distance and DoA information ($r_k$, $\theta_k$) of $K$ sources can be extracted from the received signals $\boldsymbol{Y}$. However, it's apparent from (\ref{eq:2.1}) and (\ref{eq:2.2}) that the two parameters are highly coupled, which makes conventional far-field localization algorithms fail to work in the near-field scenario.

\section{Benchmark scheme: 2D-MUSIC }
Similar to 1D-MUSIC, 2D-MUSIC leverages the orthogonality between noise and signal subspaces, while the main difference is that 2D-MUSIC utilizes near-field steering vector.  

The sample covariance matrix of the received signal $\boldsymbol{Y}$ can be obtained as:
\begin{equation}
    \boldsymbol{R} = \boldsymbol{YY}^{H}/J\approx \boldsymbol{A} \boldsymbol{R}_s \boldsymbol{A}^H + \sigma^2 \boldsymbol{I},
\label{coma}
\end{equation}
where the approximation holds when $J$ is large due to the law of large numbers. Perform eigenvalue decomposition (EVD) to $\boldsymbol{R}$, we have
\begin{equation}
\begin{aligned}
    \boldsymbol{R} = \boldsymbol{E}_s \boldsymbol{\Sigma}_s \boldsymbol{E}_s^H + \boldsymbol{E}_n \boldsymbol{\Sigma}_n \boldsymbol{E}_n^H, \label{EVD}
\end{aligned}
\end{equation}
where $\boldsymbol{E}_s$ and $\boldsymbol{E}_n$ represent the signal and noise subspace, respectively. After the division of the two subspaces, the 2D-MUSIC spatial spectrum can be constructed:
\begin{equation}
     P_{\text{2D-MUSIC}}(r,\theta) = \frac{1}{\boldsymbol{a}^H(r,\theta) \boldsymbol{E}_n \boldsymbol{E}_n^H \boldsymbol{a}(r,\theta)},
\end{equation}
where $\boldsymbol{a}(r,\theta)$ is the near-field steering vector with elements given by (2), $r \in [r_{\rm{\min }},r_{\rm{\min }}+\Delta_r,\dots,r_{\rm{\max }}]$ and $\theta=[\theta_{\rm{\min }},\theta_{\rm{\min }}+\Delta_\theta,\dots,\theta_{\rm{\max }}]$ are discretized range and angle values, with $\Delta_r$ and $\Delta_\theta$ representing the sampling interval of distance and angle, respectively. Thus, the total number of grids for 2D-MUSIC search is $n_r\times n_\theta$, with $n_r = (r_{\rm{\max }}-r_{\rm{\min}})/\Delta_r$ and $n_\theta = (\theta _{\rm{\max }}-\theta _{\rm{\min }})/\Delta_{\theta}$. 

For near-field sensing, the 2D-MUSIC algorithm has a total complexity of $ O[M^3+M^2J+n_\theta n_r(M-K)(M+1)]$. With the valid division of 2D grids, the value of $n_\theta n_r$ is comparable or even larger than $M^2$, thus, the overall complexity largely depends on the spectrum searching process, i.e., $O[M^4]$. When $M$ is large, the complexity is prohibitive.

\section{Proposed Algorithm}
Based on the above discussions, in order to achieve super-resolution sensing with low complexity, an effective approach is to minimize the search scope of the 2D spectral peak search. To this end, we propose a novel algorithm to firstly  eliminate those angle-distance regions that do not contain any source, thereby narrowing the angle-distance region that requires further 2D-MUSIC search. For angle domain, FFT can be used to firstly exclude angle regions without any sources, while for distance domain, the 1D beamforming is used in the selected angle regions to further refine the search scope. The specific procedures are described in the following.

$\mathbf{Step\, \, 1}$: Angle Cluster Determination Stage: In this stage, we apply FFT and IFFT with $S \geq M$ sampling points to the column and row vectors of sample covariance matrix $\boldsymbol{R}$ respectively. Then the spectral function $\boldsymbol{p}_\theta = [p_{\theta1},\dots,p_{\theta S}]$ that contains angle information of the sources can be obtained:
\begin{equation}
    \boldsymbol{p}_\theta = \text{diag}(\boldsymbol{WRW}^{-1}), \label{eq:FFT}
\end{equation}
where $\text{diag(·)}$ means extracting the diagonal elements of a matrix to form a vector, and $\boldsymbol{W} = [\boldsymbol{w}_1, \dots , \boldsymbol{w}_S]^T$ represents the DFT matrix, with $\boldsymbol{w}_s = [1, \dots, e^{\frac{2\pi j(k-1)(s-1)}{S}}, \dots ,e^{\frac{2\pi j(S-1)(s-1)}{S}}], k = 1,\dots,S$.

\begin{figure}[ht!] 
\centering
\includegraphics[width=0.75\hsize]{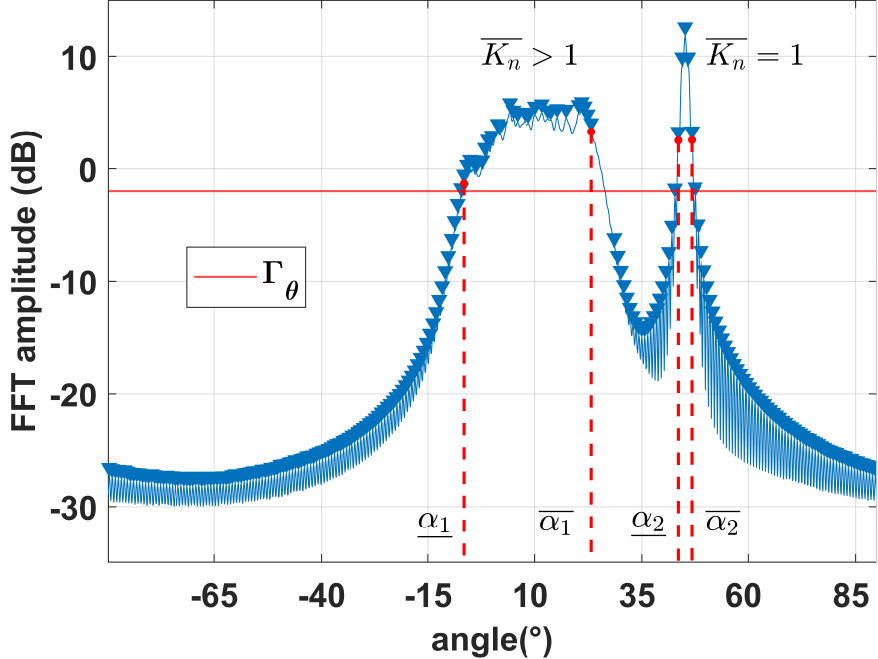} 
\caption{Illustration of how to select the spread threshold $\Gamma_{\theta}$ and divide angle clusters (adjacent red dots represent the lower/upper bounds of angle clusters respectively).}
\label{fig:3}
\end{figure}

The output of $\mathbf{Step\, \, 1}$ is illustrated in Fig.\ref{fig:3}. There may exist two different peak shapes. 
One type of peak has a relatively flat spectrum, corresponding to the signal sources that exhibit spectral spread when the FFT algorithm is used for angle search. We call it “close sources”. The other type has a sharp peak, corresponding to the sources whose angle can be determined directly using the FFT algorithm. We call it “distant sources”. Moreover, through the peak search in angle domain, regions where no signal source exists can be identified, thus allowing for the reduction of the size of angle domain that needs to be further processed. 

Due to the spectral spread phenomenon associated with close sources, the number of spectral peaks may be greater than the number of actual sources, i.e., \( \hat{K} \) $\geq$ \( K \). The set of amplitudes of these \( \hat{K} \) peaks is denoted as $\{\boldsymbol{p}_{\Theta}\}$, where $\boldsymbol{p}_{\Theta} = [p_{\Theta1},\dots, p_{\Theta i},\dots, p_{\Theta \hat{K}}] \subset \boldsymbol{p}_{\theta}$. 
To reduce the scope of the 2D search while encompassing all close sources in the angle estimation, the concept of angle clusters is defined as collective sets of angles that contain all the \( \hat{K} \) spectral peaks, denoted as $\boldsymbol{\alpha}_{n}, n=1,\dots,L$, assuming that there are $L$ angle clusters. And $\Gamma_\theta = \min(\{\boldsymbol{p}_{\Theta}\}) - \delta_\theta$ is set as the spread threshold, where $\delta_\theta$ is a parameter. After that, continuous angle samples with spectral amplitude \( \boldsymbol{p}_{\theta} >  \Gamma_\theta \) are selected to form angle clusters.
As shown in Fig.\ref{fig:3}, different angle clusters $\boldsymbol{\alpha}_{n}=[\underline{\boldsymbol{\alpha}}_{n},\overline{\boldsymbol{\alpha}}_{n}], n=1,2$ can be obtained with the valid set of $\Gamma_\theta$, where $\underline{\boldsymbol{\alpha}}_{n}$ and $\overline{\boldsymbol{\alpha}}_{n}$ are the lower and upper bounds of the angle cluster, respectively. The number of spectral peaks in $\boldsymbol{\alpha}_{n}$ is denoted as $\bar{K}_n$.

If $\bar{K}_n = 1$, there's only one candidate angle in the $n$-th angle cluster, which means that the source at that angle is distant source, and a single angle value is determined, denoted as $\theta_j$. For the case where $\bar{K}_n > 1$, there may be one or more close source whose spectral peaks are mixed together, so it's hard to resolve every single source. Thus, it's necessary to shrink the distance range and perform a refined 2D-MUSIC to get the precise location of the sources, which is shown in following steps.

$\mathbf{Step\, \, 2}$: Distance Cluster Determination Stage: If $\bar{K}_n = 1$, since there is no spectrum spread issue, high-precision distance estimation can be conducted directly. Based on $\theta_j$ obtained from the previous step, a 1D-MUSIC function with $n_r$ sampling points in distance domain is constructed:
\begin{equation}
    P_{MUSIC}(r)=\frac{1}{\boldsymbol{a}(r,\theta_j)^H\boldsymbol{E}_n\boldsymbol{E}_n^H\boldsymbol{a}(r,\theta_j)},\label{eq:3.3}
\end{equation}
where $\boldsymbol{a}(r,\theta_j)$ is the near-field steering vector with the fixed angle $\theta_j$ and $r \in [r_{\rm{\min }},r_{\rm{\min }}+\Delta_r,\dots,r_{\rm{\max }}]$, $\boldsymbol{E}_n$ represents the noise subspace obtained by (\ref{EVD}). By searching for spectral peaks from the $n_r$ sampling points, the corresponding distance value of the source $r_j$ is obtained.

For cases when $\bar{K}_n > 1$, for each angle cluster $\alpha_n = [\underline \alpha_n, \overline \alpha_n  ]$, we construct the following beamforming algorithm in terms of distance parameter with $n_r$ sampling points:
\begin{align}
    P_{low}(r)&=\boldsymbol{a}^H(r,\underline {\alpha}_n)\boldsymbol{Ra}(r,\underline {\alpha}_n), \notag \\ 
    P_{up}(r)&=\boldsymbol{a}^H(r,\overline {\alpha}_n)\boldsymbol{Ra}(r,\overline {\alpha}_n), \label{eq:3.4}
\end{align}
where $\boldsymbol{a}(r,\underline {\alpha}_n)$ and $\boldsymbol{a}(r,\overline {\alpha}_n)$ are the near-field steering vectors on the direction of the two bounds of the angle cluster.

\begin{figure}[ht!] 
\centering
\includegraphics[width=0.8\hsize]{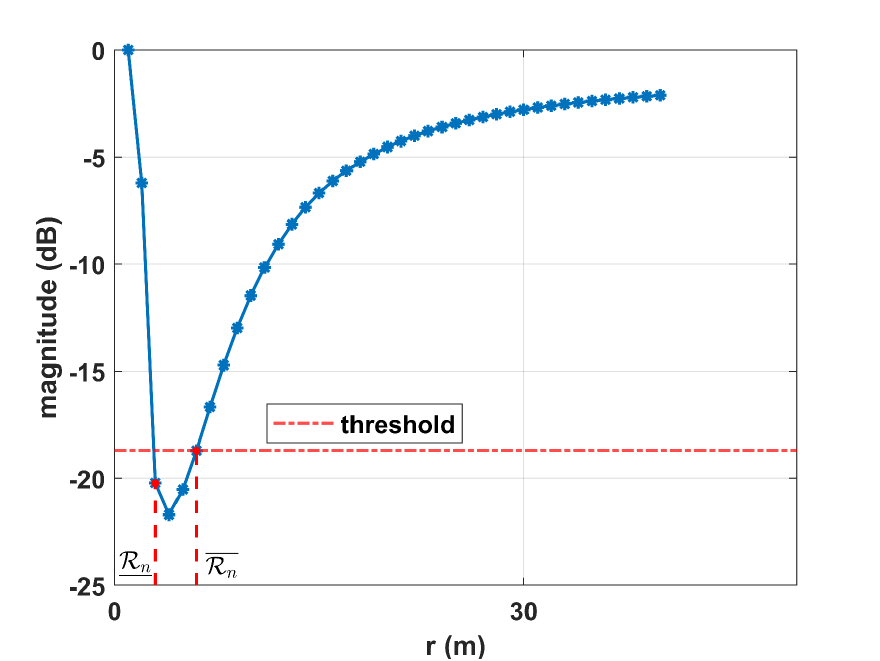}
\caption{Illustration of how to determine distance clusters $\mathcal{R}$, where $\underline{\mathcal{R}_n}$ and $\overline{\mathcal{R}_n}$ represent the lower and upper bound of $\mathcal{R}_n$.}
\label{fig:4}
\end{figure}

Assuming that $\underline{\boldsymbol{p}}_{r}$ and $\overline{\boldsymbol{p}}_{r}$ represent the set that contains the spectral amplitude of the $n_r$ sampling points of $P_{low}(r)$ and $P_{up}(r)$, respectively. Due to the characteristic of near-field beamforming, which spreads energy from the true angle-distance values  $(r_k,\theta_k)$, and because the search angles $\underline{\boldsymbol{\alpha}}_n$ and $\overline{\boldsymbol{\alpha}}_n$ are on either side of the true value $\theta_k$, the corresponding spectral peak pattern will exhibit local minima, as shown in Fig.\ref{fig:4}. Therefore, if we can find the candidate distance set $\underline{\boldsymbol{\beta}}_n$ and $\overline{\boldsymbol{\beta}}_n$ on the direction of $\underline {\boldsymbol{\alpha}}_n$ and $\overline {\boldsymbol{\alpha}}_n$ respectively, the true distance value exists in the intersection of the two distance sets, and we call it the distance cluster, denoted as $\mathcal{R}_n$. Continuous distance samples with spectral amplitude $\underline{\boldsymbol{p}}_r \leq \min\{\underline{\boldsymbol{p}}_r\} + \delta_d$ and $\overline{\boldsymbol{p}}_r \leq \min\{\overline{\boldsymbol{p}}_r\} + \delta_d$ are selected to form $\underline{\boldsymbol{\beta}}_n$ and $\overline{\boldsymbol{\beta}}_n$, respectively, where $\delta_d$ is a parameter similar to $\delta_\theta$. The true distance value exists in the distance cluster:
\begin{equation}
    \mathcal{R}_{n} = \underline{\boldsymbol{\beta}}_n \, \cap \,\overline{\boldsymbol{\beta}}_n. \label{distance}
\end{equation}
\vspace{-2.7em}

$\mathbf{Step\, \, 3}$: Super-Resolution Sensing Stage: For angle clusters with close sources, within the angle cluster $\boldsymbol{\alpha}_n$ and the distance cluster $\mathcal{R}_{n}$, the 2D-MUSIC algorithm is used to achieve joint estimation and automatic pairing of angle and distance within that scope, in which the number of grids in angle and distance domain is $n_\theta' \ll n_\theta$ and $n_r' \ll n_r$. Assume that there are $K_{close}$ close sources. For distant sources, we directly output the angles $\theta_j$ and corresponding distance parameters $r_j$ obtained in $\mathbf{Step} \, \, \mathbf{2}$, the number of distant sources denoted as $K_{distant}$. Check whether $K_{close}+K_{distant} = K$. If so, output the angle-distance values of $K$ sources. Otherwise, it is considered that the distance cluster is not sufficient to include all signal sources, which is probably due to the signal sources within the angle cluster having large differences in distance, thus affecting each other in the spectral peak search. In this case, it is necessary to expand the distance cluster scope.

$\mathbf{Step\, \, 4}$: Within the expanded distance cluster range, 2D-MUSIC is used for spectral peak search. Thereafter, repeat $\mathbf{Step\, \, 3}$ and $\mathbf{4}$ until all signal sources are localized.

\begin{algorithm}[!ht]
    \caption{Proposed FFT-enhanced  super-resolution near-field sensing algorithm}
    \label{alg:FFT-MUSIC}
    \renewcommand{\algorithmicrequire}{\textbf{Input:}}
    \renewcommand{\algorithmicensure}{\textbf{Output:}}
    
    \begin{algorithmic}[1] 
        \Require 
        the received signal matrix $\boldsymbol{Y}$ and the number of the signal sources $K$; 
        \Ensure
        angle and distance of the signal sources;
        
        \State Calculate the sample covariance matrix $\boldsymbol{R}$ based on (\ref{coma}); 
        
        \State Perform eigenvalue decomposition to $\boldsymbol{R}$, and obtain the noise subspace matrix $\boldsymbol{E}_n$.
        
        \State Obtain the angle spectral function $\boldsymbol{p}_\theta$ based on (\ref{eq:FFT}).
        
        \State Find peaks of $\boldsymbol{p}_\theta$. Based on the spread threshold $\Gamma_\theta$, obtain angle clusters $\boldsymbol{\alpha}_n, n \in [1,\dots,L]$.
        
        \For{each $n \in [1,\dots,L]$} 
        \State calculate the number of peaks $\bar{K}_n$ in the $n_{\mathrm{th}}$ cluster.
        \If{$\bar{K}_n > 1$} 
        \State Calculate the spectral function of distance domain $P_{low}(r)$ and $P_{up}(r)$ from (\ref{eq:3.4}).
        \State Obtain the distance clusters $\mathcal{R}_n$ based on (\ref{distance}).
        \State Perform the 2D-MUSIC algorithm for peak search in the identified angle-distance domain.
        \Else 
        \State Obtain the specific angle value $\theta_j$.
        \State Estimate corresponding distance value $r_j$ using (\ref{eq:3.3}).
        \EndIf
        \EndFor
        \State Sum the number of close sources $K_{close}$ as well as distant sources $K_{close}$.
        \If{$K_{close} + K_{distant} = K$} 
        \State Output the angle-distance parameters of all the signal sources.
        \Else 
        \State Update the distance clusters and back to step 8.
        \EndIf
    \end{algorithmic}
    
\end{algorithm}
\vspace{-1.5em}

The detailed steps for the proposed low complexity near-field super-resolution algorithm are summarized in Algorithm 1. Specifically, since $\theta_k$ and $r_k$ are directly estimated instead of reckoning the intermediate variables which exploits the Taylor approximation, unlike existing algorithms like \cite{RR-MUSIC}\cite{RD-MUSIC}, the proposed algorithm in this letter does not require the array spacing to be less than $\frac{\lambda}{4}$, which is compatible with typical communication systems. 

\section{NUMERICAL RESULTS}
In this section, numerical examples are provided. The BS is assumed to be equipped with $M$ = 512 antennas with $\lambda/2$ spacing. The carrier frequency is set to be 30 GHz. Assuming that there are four sources, i.e. $K$ = 4, with angles and distances set as $[(3m,6^\circ),(4m,7^\circ),(5m,8^\circ),(32m,20^\circ)]$. Without loss of generality, assuming that $r_k \in [0m,40m]$ and $\theta_k \in [-20^\circ,40^\circ], k=1,\dots,4$ for simplification, and $\Delta_\theta = 0.1, \Delta_r = 0.2$, thus $n_\theta = 600, n_r = 200$. The snapshot $J$ is set as 200, and the number of FFT sampling points $S=1024$. 

\begin{figure}[ht!]
\centering
\subfigure[localization of distant sources]{\label{fig:subfig1}\includegraphics[width=0.48\linewidth]{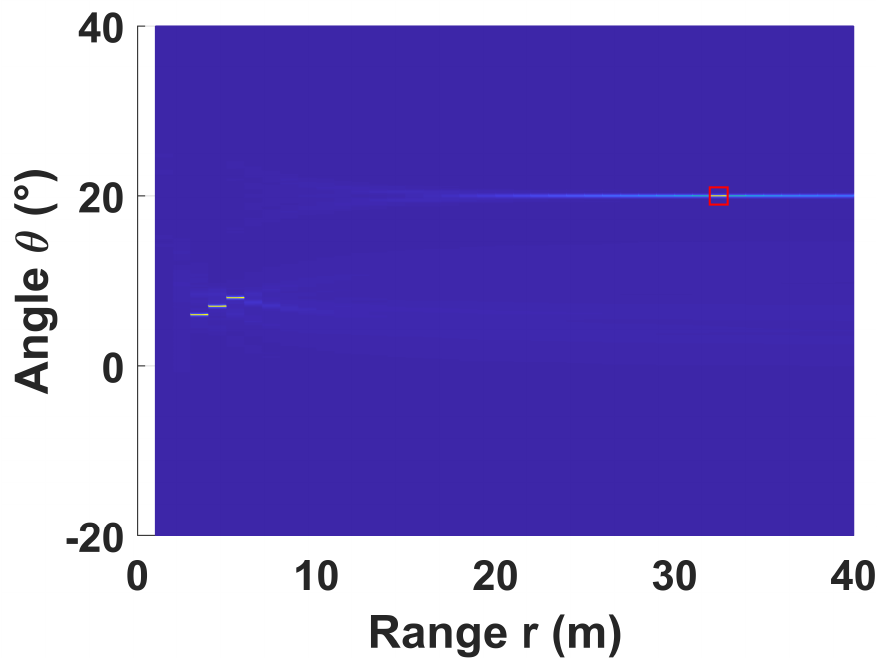}}
\subfigure[localization of close sources]{\label{fig:subfig2}\includegraphics[width=0.48\linewidth]{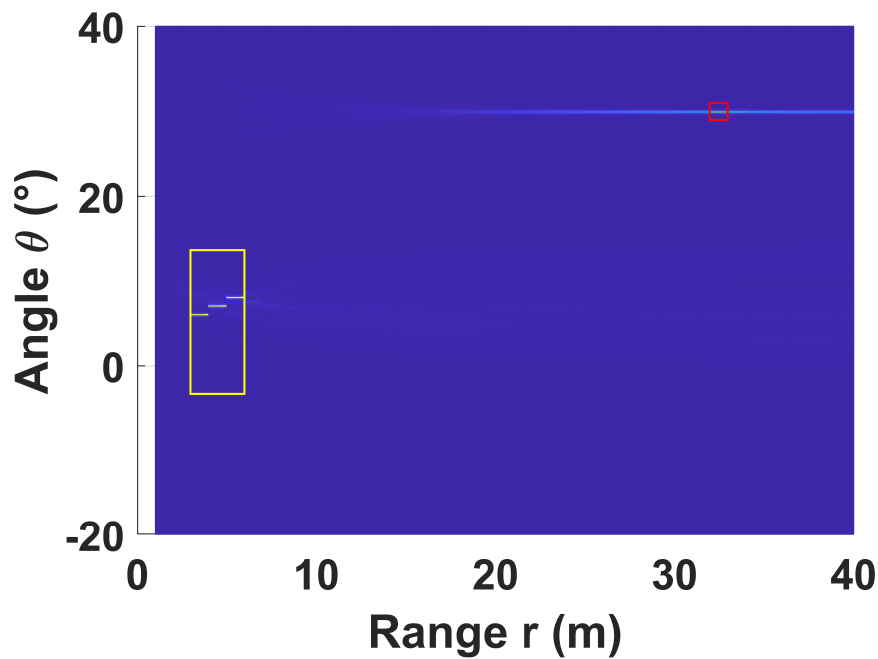}}
\caption{The performance of source localization in the near-field region.}
\label{fig:5}
\end{figure}

Fig.\ref{fig:5} illustrates the localization process of the proposed algorithm for both distant and close sources. It is evident that the proposed algorithm can achieve super-resolution localization for near-field sources, regardless of whether the sources are close or distant. To be specific, distant sources will be localized first, as shown by the red box in Fig.\ref{fig:subfig1}.Then the close sources will be estimated by the 2D-MUSIC algorithm in the refined angle-distance domain, as is shown in the yellow box in Fig.\ref{fig:subfig2}.
Since a confined 2D-MUSIC is employed for the final localization stage, the algorithm can theoretically attain the same performance as the 2D-MUSIC algorithm.
\vspace{-1.2em}
\begin{table}[h]
\caption{Running time of different near-field localization algorithms.}
\label{tableI}
\small
\setlength{\tabcolsep}{3pt}
\begin{tabular}{|c|c|c|c|c|}
\hline
algorithm & 2D-MUSIC & RR-MUSIC & RD-MUSIC & Proposed \\
\hline
running times(s) & 22.07 & 4.56 & 4.25 & 3.38 \\
\hline
\end{tabular}
\end{table}
\vspace{-0.5em}

Furthermore, the proposed algorithm has a total complexity of $O[M^3+M^2J+2MS\log_{2}(S)+2Ln_rM^2+Ln_\theta^{'} n_r^{'}(M-K)(M+1)]$. The complexity is effectively reduced by means of the FFT angle search as well as the refined 2D-MUSIC peak search, thus achieving comparable low complexity as RR-MUSIC and RD-MUSIC. The running times of different algorithms are obtained through simulations, as shown in Table \ref{tableI}.

\begin{figure}[ht!]
\centering
\subfigure[RMSE of distance versus SNR]{\label{fig: sub_figure71}\includegraphics[width=0.48\linewidth]{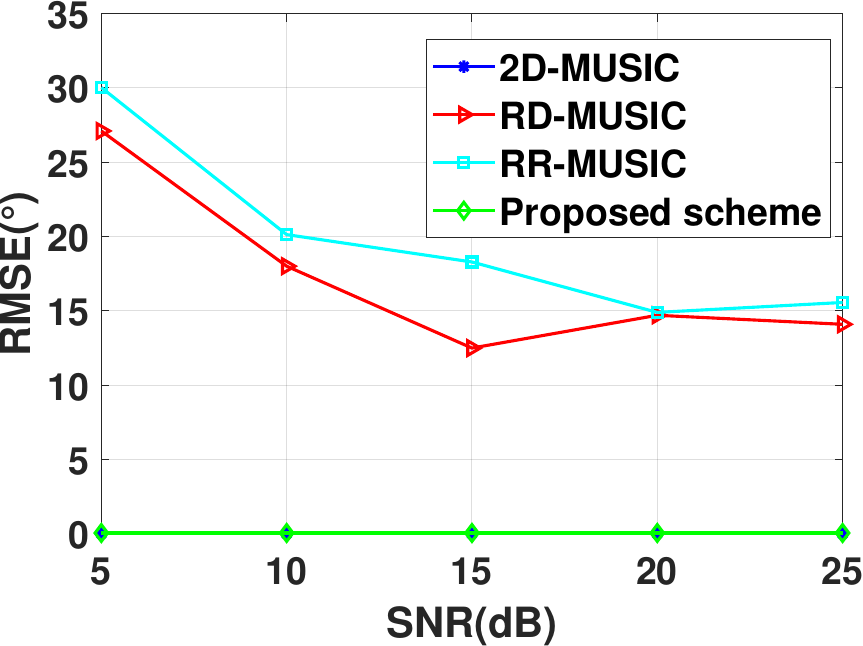}}
\subfigure[RMSE of angle versus SNR]{\label{fig: sub_figure72}\includegraphics[width=0.48\linewidth]{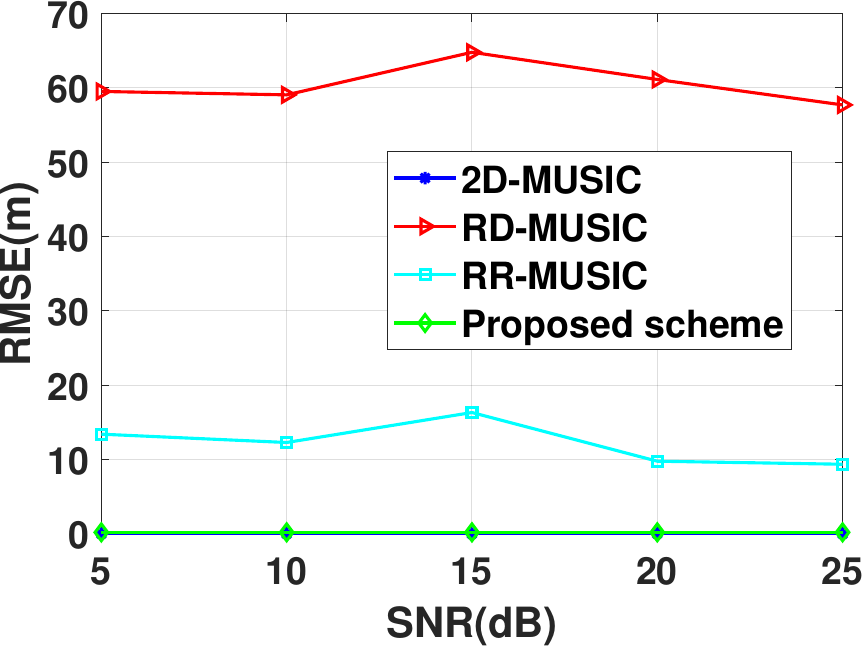}}
\caption{Comparison of localization accuracy of different near-field localization algorithms.}
\label{fig:7}
\end{figure}

The localization performance of these algorithms is shown in Fig.\ref{fig:7}. The algorithm proposed in this letter achieves localization accuracy comparable to that of 2D-MUSIC. Furthermore, it is evident that when $d = \frac{\lambda}{2}$, both RD-MUSIC and RR-MUSIC algorithm exhibit severe estimation errors. This is attributed to the fact that once $d > \frac{\lambda}{4}$, both algorithms are prone to generating numerous pseudo peaks, which in turn, compromises the estimation accuracy.  
\vspace{-1em}
\vspace{-2em}

\section{Conclusion}
This paper addressed the problem of near-field source localization. Considering the fact that existing near-field localization algorithms suffer from incompatibility with general antenna architectures or excessive computational complexity, a FFT-enhanced super-resolution near-field localization algorithm that further reduces complexity was proposed. Numerical results validated the effectiveness of the algorithm.


\end{document}